# Effect of LH and ECR waves on plasma parameters in ADITYA Upgrade tokamak

S. Aich[1,2*], S. Dolui[1,2], K. Singh[1,2], J. Ghosh[1,2], K.A. Jadeja[1,3], R.L. Tanna[1,4], K.M. Patel[1], K. Galodiya[1], A. Patel[1], L.K. Pradhan[1], B.K. Shukla[1], H. Mistry[1], J. Patel[1], H. Patel[1], D. Purohit[1], K. G. Parmar[1], P. K. Sharma[1,2], J. Kumar[1,2], B. Hegde[1,2], Abhijeet Kumar[1], Vismay Raulji[1], Praveenlal E. V.[1], T. M. Macwan[1,2], R. Kumar[1], A. Kumar[1,2], A. Kumawat[1,2], H. Raj[1], I. Hauque[1,2], Komal[1,2], S. Banerjee[1,2], P. Verma[1] and ADITYA-U team

[1]Institute for Plasma Research, Bhat, Gandhinagar 382 428
[2]Homi Bhabha National Institute, Training School Complex, Anushaktinagar, Mumbai 400 094
[3]Department of Nano Science and Advanced Materials, Saurashtra University, Rajkot 360 005
[4]Institute of Science, Nirma University, Ahmedabad 382 481

**Abstract:** The plasma discharges in ADITYA Upgrade Tokamak are produced by means of transformer action, in which Ohmically created plasma is driven by means of a secondary loop voltage. Due to reduction of plasma resistivity after a certain level of plasma temperature, Ohmic heating becomes poor and further achievement of temperature needs other heating techniques. ADITYA-U tokamak is facilitated with a 42 GHz-500 kW Electron Cyclotron Resonant Heating (ECRH) system. Also, there is a Lower Hybrid Current Drive (LHCD) system installed and operated at 3.7 GHz for driving non-inductive plasma current followed by the Ohmic current drive. Though an eventual impact in the rise of plasma temperature and plasma current due to the application of ECRH and LHCD respectively are very obvious, their energy coupling with the plasma results in several interesting outcomes in a number of experimentally measured plasma parameters. The present work addresses such impactful observations that are noticed and reported for the first time in ADITYA-U Tokamak.

## 1 Introduction

Tokamak is one of the potential candidates among the fusion devices [1-2]. In a tokamak, plasma is Ohmically driven by means of a toroidal voltage (loop voltage) that appears due to the transformer action using a set of transformer (TR) coils [2]. The isotopes of hydrogen - Deuterium and Tritium are found to be the most appropriate fuels for the nuclear fusion. To enable the D-T fusion possible, plasma temperature needs to be achieved of order of 10 keV, which is impossible to reach for an Ohmically heated plasma. Moreover, driving the plasma current for a longer duration Ohmicqally requires a cost effective Ohmic power-supply as well as an appropriate cooling mechanism in case of superconductor coils. Present fusion machines that are aimed for D-T fusion, like ITER, need other auxiliary plasma heating and current drive mechanisms to be implemented [1]. In this respect, Electron Cyclotron Resonant Heating (ECRH) mechanism is a very useful to heat the electron and eventually to raise the plasma temperature to a desired value [3-4]. On the other hand, the technique of Neutral Beam Injection (NBI) and Lower Hybrid Current Drive (LHCD) are well-established for tokamak current drive globally [5-7]. Current drive plays a crucial role for enhancing plasma duration and hence to achieve desired plasma parameters.

ADITYA Upgrade is a medium sized tokamak machine, having major and minor radii to be 75 cm and 25 cm respectively. The plasma breakdown occurs by means of an initial loop voltage ~20 volt and later the plasma is driven by ~1.8 volt at the flat-top region. The typical discharges achieve a duration of 300 ms with negative converter [8]. The tokamak is facilitated with a 42 GHz-500 kW ECRH and a 3.7 GHz-250kW LH systems. Present work discuss on the direct impact of ECRH and LH waves' launching on a number of plasma parameters. The paper is arranged as follows. Section 2 gives a very short overview of ECRH and LH systems in ADITYA-U. Then, section 3 summarizes the results that are found due to the application of ECRH and LH systems. Finally, section 4 provides a rigorous discussion and hence conclusions therefrom, followed by the future scope of work.

## 2 ECRH and LHCD systems in ADITYA-U

The Lower Hybrid system is connected with ADITYA-U tokamak at port number 5 and it is operational according to the experimental requirements [9-10]. The lower hybrid waves (LHWs) at 3.7 GHz are launched from low field side (LFS), employing passive active multi-junction (PAM) launcher [9-10]. The time for launching is chosen either at plasma flat-top or at the end of flat-top, according to the experimental requirements. On the other hand, ECRH system is connected to ADITYA-U tokamak at port 14 and launched into the plasma from LFS in O-mode. ECRH wave of 42 GHz is launched into the Ohmic plasma either to support pre-ionization or to heat plasma or for both purposes [3-4]. The details of these techniques and results will be separately discussed in future publications.

## 3 Results: Effect of plasma heating and current drive

The analysis of present paper is restricted by similar types of discharges, in which plasma is Ohmically generated and driven with loop voltage. Being the secondary of the transformer, the effective loop voltage for breakdown and current drive is typically ~ 18 V and 2 V respectively. Additional ECRH power and LHCD power are launched separately in a same or in different plasma discharges to enhance the energy and duration of plasma respectively.

### a. Diamagnetism, plasma beta and stored energy

A charged particle, gyrating about a magnetic field, shows diamagnetism by opposing the source magnetic field and so diamagnetism is an in-built nature of a tokamak plasma [2]. The resultant diamagnetic current in a tokamak is strongly dependent on the pressure gradient ($\Delta p$) and hence on the plasma density ($n_e$) and temperature ($T_e$) [11-12]. In an Ohmically driven low $p$ plasma, the diamagnetic current is insufficient to reduce the value of toroidal magnetic field $B_\Phi$ inside plasma. In contrary, the helical current gives rise to the enhancement of $B_\Phi$ in presence of plasma current and so the plasma behaves like paramagnetic element, leaving the poloidal beta ($\beta_\theta$) less than 1.0 [12].

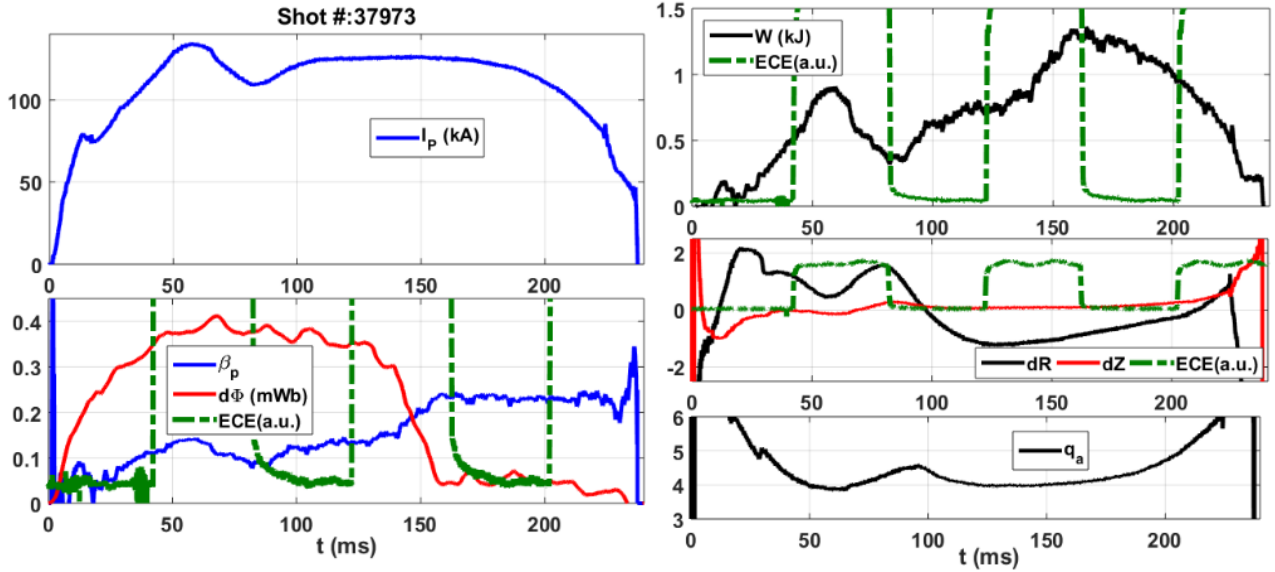

Fig 1. Time profiles of different plasma parameters, as indicated in legends, along with the launched ECRH powers (in green) for plasma discharge 37973.

During the plasma discharges in ADITYA-U tokamak, the difference in toroidal flux in presence of plasma from that of the vacuum flux i.e., diamagnetic flux $\Phi_{dia} = (\Phi_{Pl} - \Phi_{Vac})$ is measured using a diamagnetic loop system [13]. This $\Phi_{dia}$ is found to be ~ + 0.3 mWb. The positive sign of estimated $\Phi_{dia}$ essentially indicates that the plasma remains paramagnetic in this Ohmic operational regime. But, a successful coupling of ECR wave with the plasma essentially enhances $\Delta p$ and diamagnetic current is enhanced. As a final outcome, a drastic reduction in $\Phi_{dia}$ is noticed, as given in the left bottom plot of Fig 1 in red curve for a typical plasma discharge. Another typical discharge is given in Fig 2 and similar phenomena is noticed. In the plot, ECR launched signal is indicated in green. The poloidal beta $\beta_\theta$ can hence be estimated using:

$$\beta_\theta = 1 - \frac{8\pi B_\Phi}{(\mu_0 I_p)^2} \Phi_{dia}$$

where $I_p$ stands for plasma current and $\mu_0$ is free-space permeability. Time profile of $\beta_\theta$ is also shown in the same – sub-plots of Fig 1 and Fig 2 in blue curve. It clearly shows that the enhancement of $\beta_\theta$ modifies the $\Delta p$ in such a way that the diamagnetic current is enhanced and essentially the property of Ohmic plasma approaches from a paramagnetic to a diamagnetic nature. Moreover, as a result of enhancement in $\beta_\theta (\propto n_e T_e)$, the total kinetic energy of plasma also increases, as indicated by the top-right subplot in Fig 1 and middle-right subplot in Fig 2.

The introduction of LH power into the plasma discharge is found not to affect the diamagnetic property in detectable range in the present operational regime.

### b. Safety factor

ADITYA-U Ohmic plasma is operated typically with an edge safety factor of $q_a \geq 4.0$ [8] and with the maximum achieved $I_p \approx 210$ kA $q_a$ reduces to 2.4. During experiments with LH power launching, the driving loop voltage at plasma flat-top is kept low to resolute a current drive by the LHCD. According to the coupling of LH power with plasma, there is no significant change in $q_a$ is noticed. But interestingly, coupling of ECR wave with plasma is found strongly dependent on $q_a$ and a successful coupling of the Ohmic plasma is preferred at $q_a \approx 4.0$. For example, in case of plasma shot 37973 in Fig 1, plasma current is stable from 100 ms to 170 ms and one of the ECR pulses is launched in this time domain (at around 120 ms) and there is a successful coupling of the launched power with plasma during $q_a$ approaching 4.0, in contrary to the first ECR pulse, that is distinctly captured by the sudden drop in $\Phi_{dia}$. Time variation of $q_a$ is shown at the right-bottom plot in Fig 1 and 2. The similar scenario for another typical plasma discharge 37976 is shown in Fig 2.

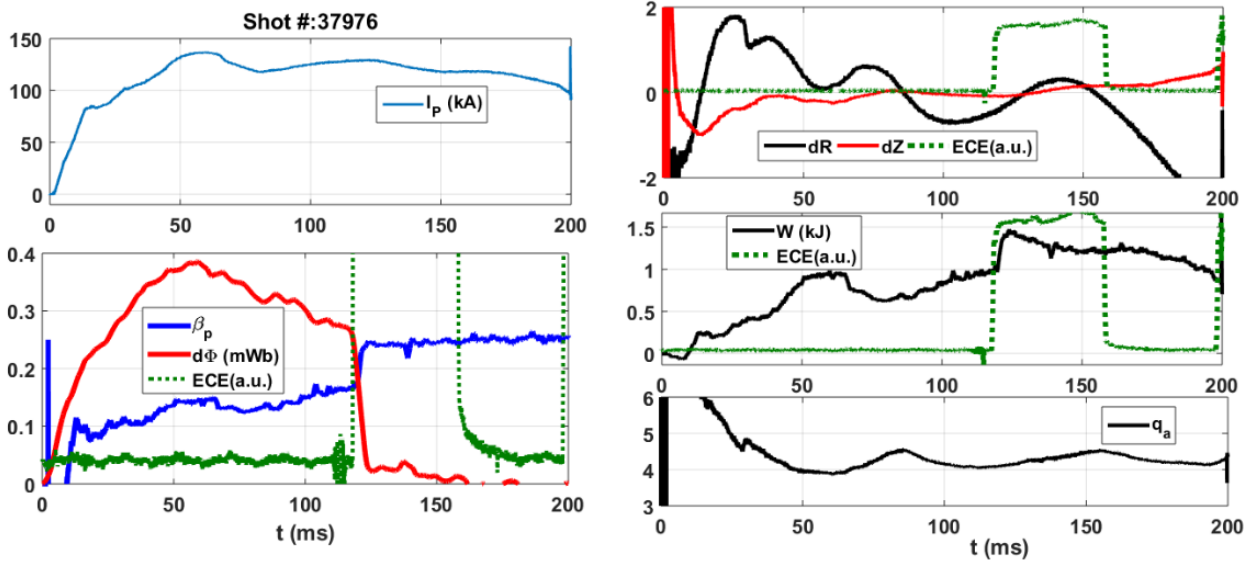

Fig 2. Time profiles of different plasma parameters, as indicated in legends, along with the launched ECRH powers (in green) for plasma discharge 37976.

### c. Plasma position

ADITYA-U plasma operation in limiter configuration is facilitated with a real-time horizontal position control system [8] and the horizontal along with the vertical movement of plasma is estimated using magnetic as well as optical signals. The plots in red and black in the middle-right panel of Fig 1 indicates vertical and horizontal position of plasma respectively. It is obvious from the plots that an ECR coupling with the plasma pushes towards low field side (outboard side). This has been observed for a number of plasma discharges and is shown for another typical discharge in Fig 2.

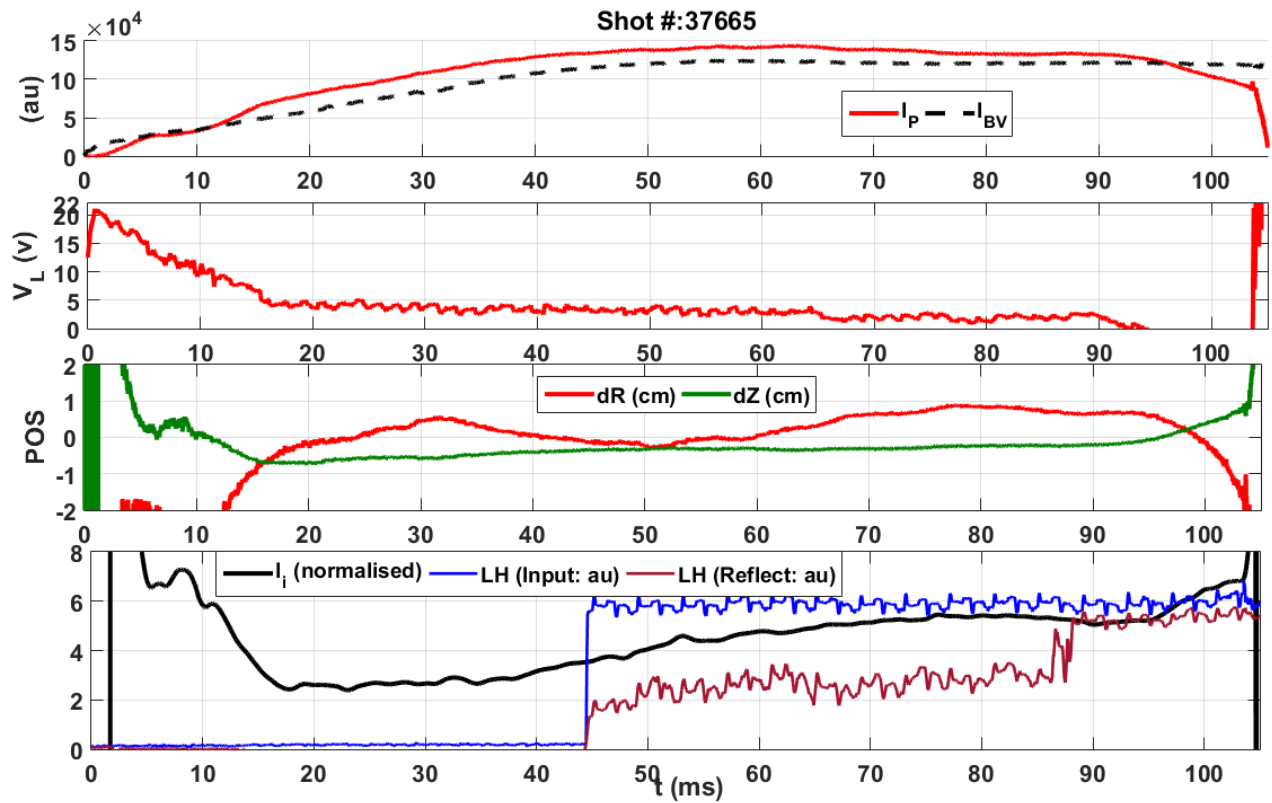

Fig 3. Time profiles of different plasma parameters, as indicated in legends, loop voltage, along with the launched (in blue) and reflected (in grey) LH powers for plasma discharge 37665.

In contrary to the behavior of horizontal movement of plasma in presence of EC wave, LH coupling is very much dependent on the positional movement towards high field side (inboard side). It is observed that an outboard position of the column is preferred for a better coupling of LH launched power. For example, Fig 3 summarizes the effect of LHCD power coupling on positional shifts and inductance (normalized) of plasma. In the topmost panel, plasma current and current through vertical magnetic field coils are summarized, whereas, second and third panels provide time variations of loop voltage and plasma positions respectively. Fourth panel summarizes the time profile of launched (blue) and reflected (grey) LH power, along with inductance of plasma column, which is estimated with the help of measured vertical magnetic field and poloidal beta, in black. The similar outcomes for another typical plasma discharge 37733 are summarized in Fig 4.

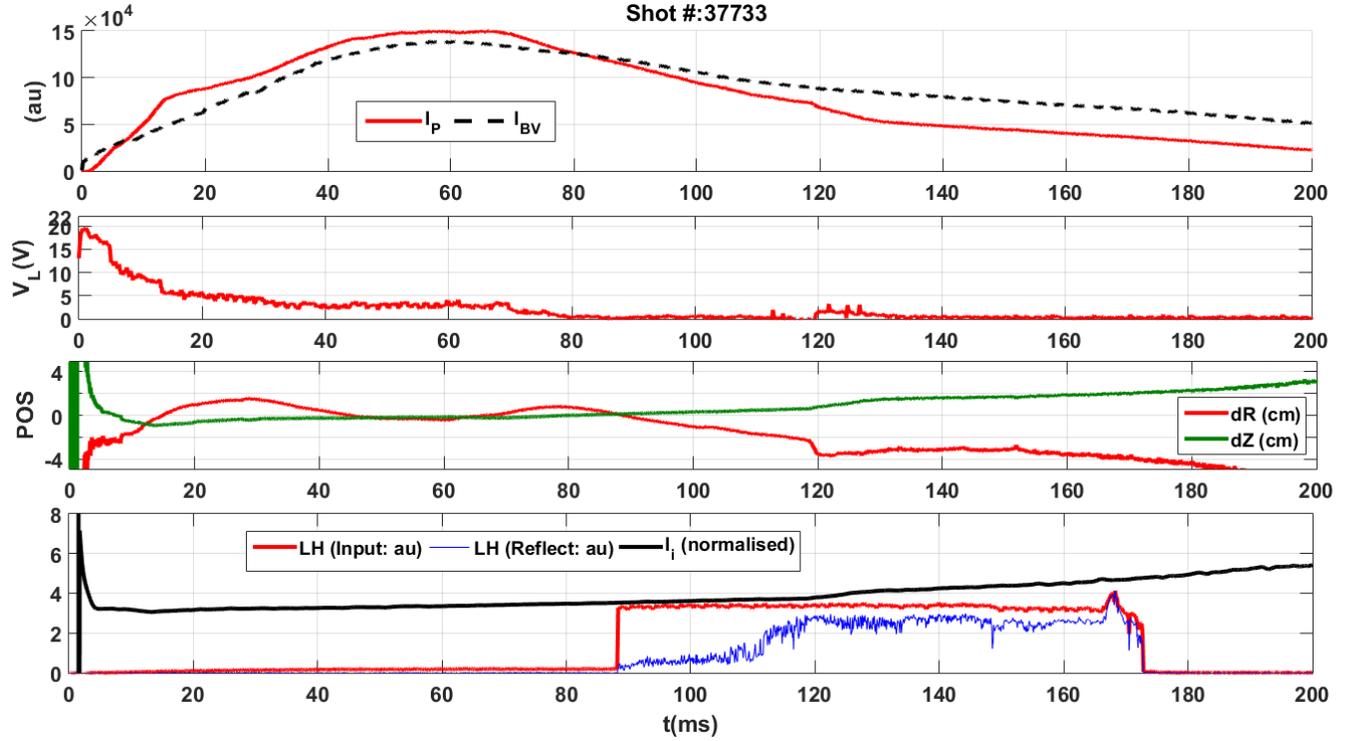

Fig 4. Time profiles of different plasma parameters, as indicated in legends, loop voltage, along with the launched (in red) and reflected (in blue) LH powers for plasma discharge 37733.

**4 Discussion and Conclusion**

The impact of current drive or plasma heating by means auxiliary mechanisms on plasma parameters is not always very straight forward to follow and thus it remains an interesting regime to be explored. A measure of the direct impact of these mechanisms on plasma parameters is not only interesting, but also helps to figure out the effective coupling of the launched power with plasma at the operated parameter domain. The present work summarizes several primitive observations on the immediate effects of LHCD and ECRH on a number of measured plasma parameters in ADITYA-U tokamak.

A satisfactory coupling of ECRH power with plasma enhances diamagnetic current, $\beta_p$ and kinetic energy of plasma. Eventually, the vertical magnetic field, required for horizontal positional equilibrium, becomes insufficient for the present plasma current and so the entire column has a tendency to move towards outboard. Also, the coupling of ECR power with the plasma occurs above some critical value of $\beta_p$ and the value is found to be 0.16 for ADITYA-U Ohmic plasma. It has also been observed that the ECRH power is coupled when $q_a \sim 4.0$, which indicates that the current density profile has a very crucial impact of the power coupling.

LHCD power coupling in ADITYA-U tokamak is noticed by comparing the launched signal and the corresponding reflected signal. It is observed that the coupling is favored when plasma remains towards outboard side and gradually gets weakened as the plasma moves towards inboard. Though this fact is being studied in more details and still cannot be explained fyllu, it will be addressed in future publications. Also, a weaker coupling of LH power with plasma results in restoring of peaked plasma current density profile that is typically noticed in Ohmically driven plasma, from a flatter one [14]. It is distinctly indicated by the rise of $l_i$ after 120 ms in discharge 37733 (Fig 4), for example, followed by a reduced LH coupling. The effect of LHCD on $l_i$ is more evident from Fig 3. Initially around ~ 45 ms, the LHCD coupling is very good and LHCD power supports off-axis current drive, thereby a rise in $l_i$ is not significant and towards 88 ms $l_i$ tries to decrease followed by a flat-top. Later (beyond 90ms) the coupling detoriates thereby reducing the current drive effect due to LHWs and thus $l_i$ tends to increase rapidly, indicating a peaking of current profile.

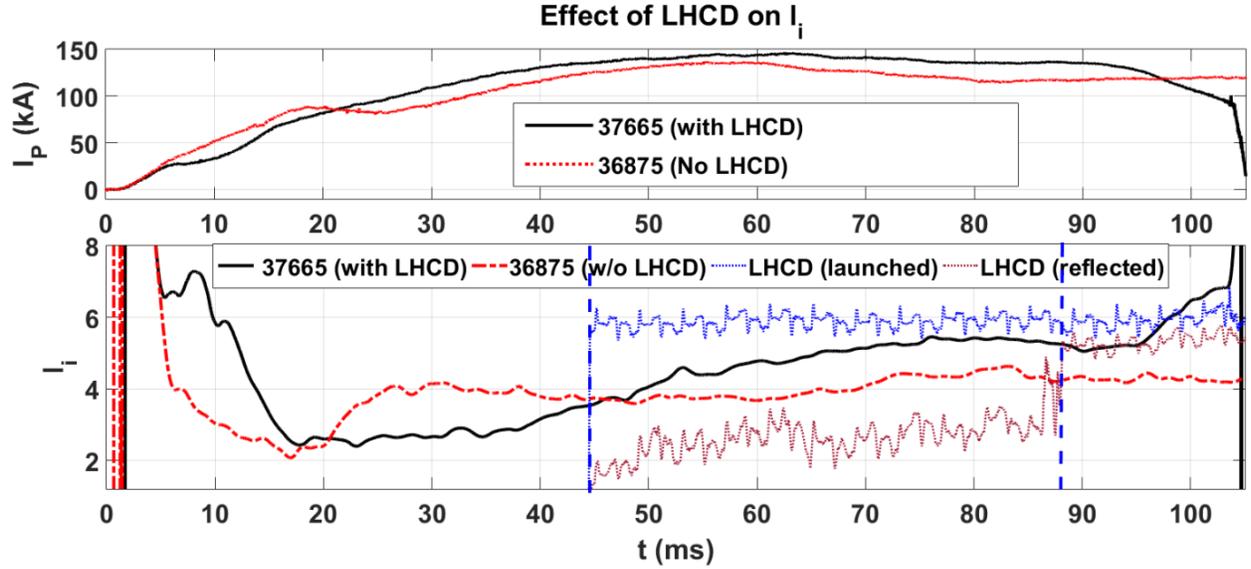

Fig 5. Time profiles of different plasma parameters, as indicated in legends, loop voltage, along with the launched (in red) and reflected (in blue) LH powers for plasma discharge 37733.

Fig 5 is prepared for a fair comparison between two plasma discharges of almost same parameters, but one (37665) with LHCD and other (35875) without LHWs. $l_i$ is strongly dependent on current density profile and a low value of $l_i$ indicates a broader profile. During plasma equilibrium, the current profile does not vary abruptly and so $l_i$ maintains almost a plateau in the plasma flat-top, as shown by shot 36875. It is very obvious from this comparison of $l_i$ that an abrupt rise in $l_i$ is due to peaking of current profile in 37665 after reduced coupling of LH power with plasma.

The entire paper summarizes primitive effect of ECRH and LHCD on a bunch of plasma parameters, though, more versatile effects in other plasma parameters along with their physical interpretations, importance are also being studied and will reported through upcoming publications in near future.

**Acknowledgemets**

The authors thank the entire team from Electronics and data acquisition divisions, who have directly or indirectly contributed in this work. Also, the authors are thankful to the entire LHCD, ECRH and AOD team members for all their supports.

______________________
*Corresponding author*
 e mail: suman.aich@ipr.res.in (Suman Aich)